\documentclass[notitlepage,superscriptaddress,aps,amsmath,amssymb,floatfix,reprint,nobalancelastpage,raggedbottom]{revtex4-1}
\usepackage[pdftex]{graphicx}
\usepackage{amsmath}
\usepackage{lipsum}
\usepackage{color}
\usepackage{soul}
\usepackage{balance}

\begin{document}
\title{Voltage-driven Building Block for Hardware Belief Networks}
\author{Orchi Hassan}
\affiliation{School of Electrical and Computer Engineering, Purdue University, West Lafayette, IN 47907, USA}
\author{Kerem Y. Camsari}
\affiliation{School of Electrical and Computer Engineering, Purdue University, West Lafayette, IN 47907, USA}
\author{Supriyo Datta}
\affiliation{School of Electrical and Computer Engineering, Purdue University, West Lafayette, IN 47907, USA}

\begin{abstract}

Probabilistic spin logic (PSL) based on networks of binary stochastic neurons (or $p$-bits) has been shown to provide a viable framework for many functionalities including Ising computing, Bayesian inference, invertible Boolean logic and image recognition. This paper presents a hardware building block for the PSL architecture, consisting of an embedded MTJ and a capacitive voltage adder of the type used in neuMOS. We use SPICE simulations to show how identical copies of these building blocks (or weighted $p$-bits) can be interconnected with wires to design and solve a small instance of the NP-complete Subset Sum Problem fully in hardware.\\ \textbf{Keywords -} Probabilistic computing, Embedded MTJ, $p$-bits, $p$-circuits, Invertible Boolean logic, Subset Sum Problem
\end{abstract}

\maketitle
\section{Introduction}

Probabilistic spin logic (PSL) has been shown to  provide a viable framework for Ising computing \cite{sutton2017intrinsic, behin2016building,shim2017}, Bayesian inference \cite{behin2016building}, invertible Boolean logic  \cite{camsari2017stochastic}, and image recognition \cite{zand2017r}. The PSL model is defined by two equations  \cite{camsari2017stochastic} loosely analogous to a neuron and a synapse. The former  is what we call the $p$-bit whose output $m_i$ is related to its dimensionless input $I_i$ by the relation

\begin{subequations}
\begin{equation}
{m_i}(t+\Delta t) = {\rm{sgn}}\{ \mathrm{rand(-1,1)} + \mathrm{tanh}({I_i}(t))\} 
\end{equation}

\noindent where rand($-$1,+1) is a random number uniformly distributed between $-$1 and +1, and $t$ is the normalized time unit. The synapse generates the input $I_i$ from a weighted sum of the states of other $p$-bits according to the relation
\begin{equation}
{I_i}(t) = I_{0} \bigg( h_i(t)+\sum_j{J_{ij}m_j} \bigg)
\end{equation}
\label{eq:PSL}
\end{subequations}

\noindent where, $h_i$ is the on-site bias and $J_{ij}$ is the weight of the coupling from $j^{th}$ $p$-bit to $i^{th}$ $p$-bit and $I_0$ is a dimensionless constant. These two equations constitute the behavioral model of PSL. The objective of this paper is to present a voltage-driven hardware building block using present day device technologies such as embedded MRAM \cite{lin200945nm} and Floating-Gate MOS transistors, such that identical copies of the same block can be interconnected with wires to implement Eqs.~\ref{eq:PSL}. \\

The paper is organized as follows: We first show a complete hardware mapping for the weighted $p$-bit by augmenting a recently introduced Magnetoresistive Random Access Memory (MRAM) type stochastic unit \cite{camsari2017implementing} with a floating gate MOS-based capacitive network \cite{ohmi1992neumos}. We then show how the results of a fully interconnected $^W\hspace{-4pt}p$-bit circuit closely approximate the the ideal equations using an example of an ``invertible'' Full Adder that can perform 1-bit addition and subtraction. Finally, we show how such invertible Full Adders can be interconnected to solve a simple instance of the NP-complete Subset Sum Problem. \\

Each example in this paper has been obtained using full SPICE models which simply uses transistors, capacitors and resistors without any additional complex circuitry or processing.

\section{Building block}

\begin{figure}[!t]
\centering
\includegraphics[width=0.98\linewidth]{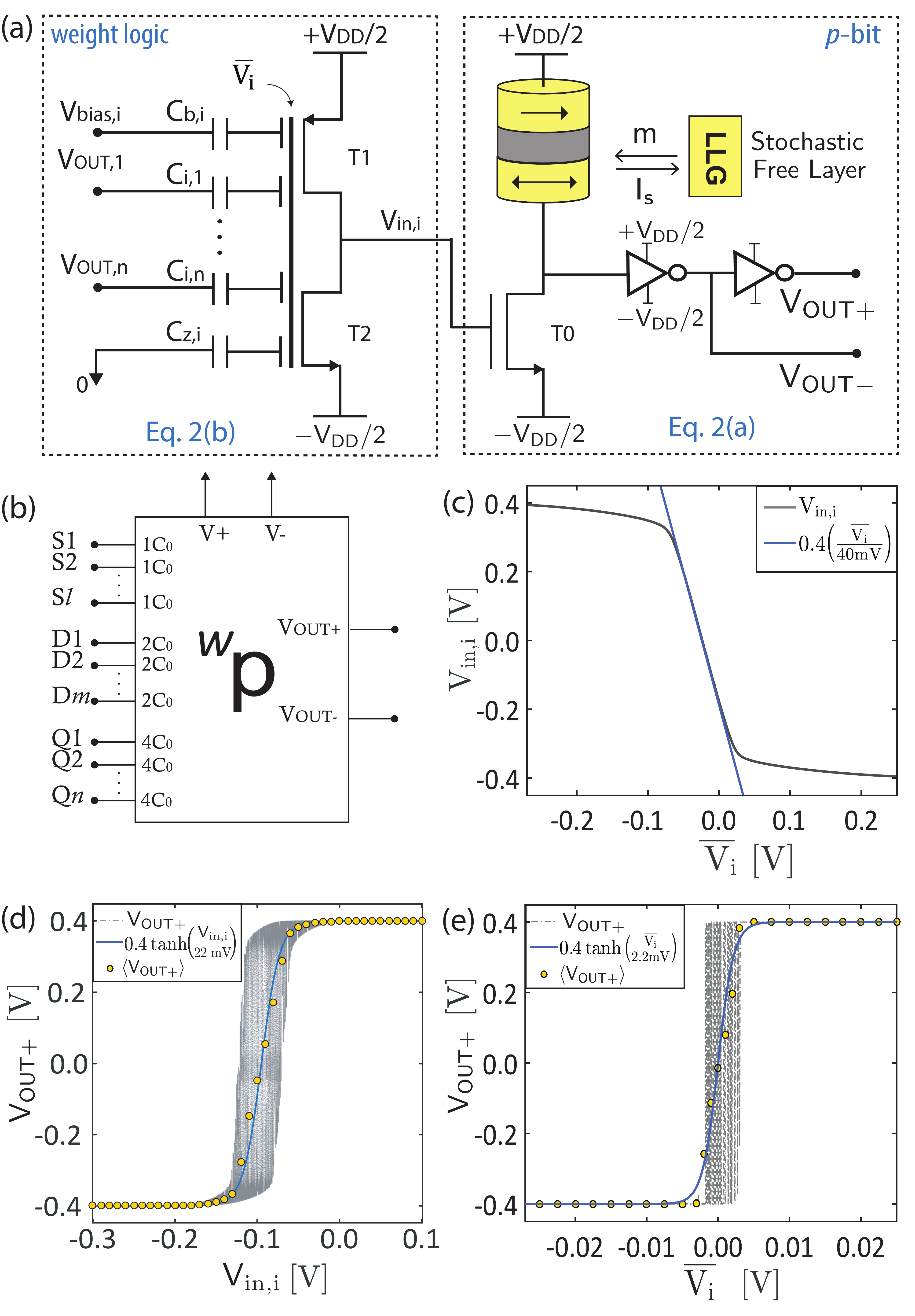}
 \caption{(a) \textbf{Voltage-driven building block} has two components corresponding to Eqs.~\ref{eq:circuit}a,b. The first is the $p$-bit implemented through an embedded low-barrier unstable MTJ \cite{camsari2017stochastic} with two inverters added to give positive and negative outputs. The low-barrier MTJ can be designed using low barrier or circular nanomagnets. The second is the capacitive voltage adder with an inverter structure on the left similar to the floating gate MOS transistors used in neuMOS devices \cite{ohmi1992neumos}. We call this combination of $p$-bit and its weight logic a weighted $p$-bit ($^W\hspace{-4pt}p$-bit). (b)Shows the the block diagram of $^W\hspace{-4pt}p$-bit. (c) Shows how an inverter helps amplify the input ($\overline{V_i}$) of the capacitive network to give $V_{in,i}$ at the gate of the $p$-bit's NMOS transistor $\rm{T0}$. (d) Shows the relation of the input gate voltage of the NMOS ($V_{in,i}$) to output ($V^{+}_{OUT}$). (e) Shows the transfer characteristics of the $^W\hspace{-4pt}p$-bit as a whole. The inputs in each case is swept from $-$0.4V to $+$0.4V in 1 $\mu$s. The yellow dots are time averaged values at each point over 300 ns and the solid blue lines are numerical fits. The magnet used in the simulations is defined by parameters in\cite{camsari2017implementing}: $M_s=1100emu/cc,D=22nm,t=2nm,\alpha=0.01$. All transistors were modeled using minimum size (nfin=1) 14 nm HP-FinFET Predictive Technology Models with $\rm{V_{DD}=0.8V}$ and $\rm{T=300K}$.}
\label{fi:fig1}
\end{figure}
%%%%%%%%%%%%%%%%%%%%%%%%%%

Our building block has two components corresponding to the two Eqs.~\ref{eq:PSL}a,b. Eq.~\ref{eq:PSL}a is implemented by the $p$-bit in Fig.\ref{fi:fig1}a which consists of an embedded low-barrier unstable MTJ coupled to two CMOS inverters which provides a stochastic output whose average value is controlled by the input voltage:
\begin{subequations}
\begin{equation}
V_{out,i}= \frac{V_{DD}}{2}  \mathrm{sgn} \left(\mathrm{rand}(-1,+1) + \mathrm{tanh} \frac{V_{in,i}}{V_{0}} \right) 
\end{equation}
\noindent where $\pm V_{DD}/2$ are the supply voltages, and $V_0$ is a parameter ($\sim22 \rm \ mV$) describing the width of the sigmoidal response.

%% add description of the MTJ-NMOS unit and how G0 is found from Notes somewhere here

The value of $V_0$ depends on the details of the 1T/1MTJ  in the embedded MRAM structure \cite{camsari2017implementing} and the transistor characteristics. The conductance, $G_0$ of the MTJ is chosen to match the MTJ switching characteristics to the transistors in the $^W\hspace{-4pt}p$-bit so that the overall transfer characteristics is centered at zero as shown in Fig.~\ref{fi:fig1}e. To do that, an input voltage of $\rm{\overline{V_i}}=$0V is applied at the input of T1 and T2 transistors turning both of them ON ($\rm{\vert V_{GS} \vert = 0.4V}$) and $G_0$ is swept to observe the outputs. The $G_0$ value for which $\rm{V^{+}_{OUT}}$=$\rm{V^{-}_{OUT}}=0V$ is the value chosen to be the MTJ conductance. For minimum sized 14nm HP-FinFET transistors models with $\rm{V_{DD}=0.8V}$, $1/G_0\approx$ 62 k$\Omega$  and it seems reasonable considering the RA-products of modern MTJs \cite{mizrahi2018neural}.

Eqs.~\ref{eq:PSL}b is implemented by the weighted synapse portion of Fig.~\ref{fi:fig1}a , which is a capacitive voltage adder just like those used in neuMOS devices \cite{ohmi1992neumos,nakamura2015neuron}. We can write
\begin{equation}
\overline{V}_{i}= \frac{V_{bias,i} C_{b,i} + \sum_j{V_{out,j} C_{ij}}}{C_g+C_{z,i}+C_{b,i} + \sum_j{C_{ij}}}
\end{equation}

\noindent Note that the capacitive voltage divider typically attenuates the voltage $\overline{V}_{i}$ at its output, and the inverter scales it up to $V_{in,i}$ as shown in Fig.~\ref{fi:fig1}c, the two being related approximately by 
\begin{equation*}
V_{in,i} \approx \frac{V_{DD}}{2}  \mathrm{tanh} \frac{\overline{V}_{i}}{\nu_0}  \ \ \ \ \ \ \ \ \  \ \ \ \ \ \ \ \ \ \
\end{equation*}
\vspace{-0.2in}
\begin{equation}
\approx \frac{V_{DD}}{2 \nu_0} \ \overline{V}_{i} \ \ \   \mathrm{if} \ \ \  \overline{V}_{i} \ll \nu_0 
\end{equation}
\label{eq:circuit}
\end{subequations}

\noindent where $\nu_{0}$ is a parameter characteristic of the inverter. Eqs.~\ref{eq:circuit}a,b can be mapped onto the PSL Eqs.~\ref{eq:PSL}a,b by defining
\begin{subequations}
\begin{equation}
m_i = \frac{V_{out,i}}{V_{DD}/2}, \ \  I_i = \frac{V_{in,i}}{V_0}
\end{equation}
\begin{equation}
C_{b,i} = b_{i} C_{0} \ \  C_{z,i} = z_{i} C_{0}
\end{equation}
\begin{equation}
h_i = b_{i} \frac{V_{bias,i}}{V_{DD}/2}, \ \ J_{ij} = \frac{C_{ij}}{C_{0}}
\end{equation}
\begin{equation}
I_0 = \frac{(V_{DD}/2\nu_{0})(V_{DD}/{2V_0)}}{(C_g/C_0)+z_{i}+b_{i} + \sum_j{J_{ij}}} 
\end{equation}
\label{eq:map}
\end{subequations}

\noindent$C_g$ is the intrinsic gate capacitance of the neuMOS inverter. The significance of $C_{0}$ is that we assume the input is composed of many identical capacitors $C_{0}$, and that the weights $J_{ij}$ have been designed to have \emph{integer} values such that $C_{ij}$ can be implemented by connecting $J_{ij}$ elementary capacitors in parallel. The other coefficients $z_{i}$, $b_{i}$ are also integers. We adjust the number $b_i$ of bias capacitors to facilitate external biasing and the number $z_{i}$ of grounded capacitors to make $z_{i}+b_{i} + \sum_j{J_{ij}} = K$ a constant, so that $I_0$ is independent of index $i$:
\begin{equation}
I_0 = \frac{(V_{DD}/2\nu_{0})(V_{DD}/2V_0)}{(C_g/C_0)+K}
\label{eq:io} 
\end{equation}

Note that $K$ is usually a fairly large number equal to the sum of all the weights, and to implement an $I_0\sim \ 1$ it is important to keep the factor $(V_{DD}/2\nu_{0})(V_{DD}/2V_0)$ to be much greater than 1. This is the reason for using an inverter between the capacitive voltage adder and the $p$-bit. Our model neglects any leakage resistances associated with the capacitive weights. Modern transistors with thin oxides can have gate leakage currents $\sim$1nA, with RC $\sim\mu$s-ms. This should not affect the weighting, since the examples presented here operate at sub-ns time scales. For slower neurons, it may be advisable to use thicker oxides for the capacitive weights to ensure lower leakage.

Fig.~\ref{fi:fig1}b shows the icon we use to represent our building block which we call a weighted $p$-bit. The input consists of three types of inputs designated S, D and Q having capacitances $C_0$, $2 \ C_0$ and $4 \ C_0$. Combinations of these are used to implement different weights 
$J_{ij}$ and different bias $h_{i}$. Each block has two outputs $V_{OUT}^{+}, V_{OUT}^{-}$. The choice of output depends on the sign of the corresponding $J_{ij}$. Similarly different signs of $h_{i}$ are implemented by choosing $V_{bias,i}$ to be $+V_{DD}/2$ or $-V_{DD}/2$.

\section{Invertible full adder}
In PSL, any given truth table can be implemented using Eq.~\ref{eq:PSL} by choosing an appropriate $[J]$ and $[h]$ matrices \cite{camsari2017stochastic}. Here we show how those $[J]$ and $[h]$ are mapped onto physical hardware using our proposed building block using only transistors, resistors and capacitances. 

%%%%%%%%%%%%%%%%%%%%%%%%%%
\begin{figure}[!t]
\centering
\includegraphics[width=0.95\linewidth]{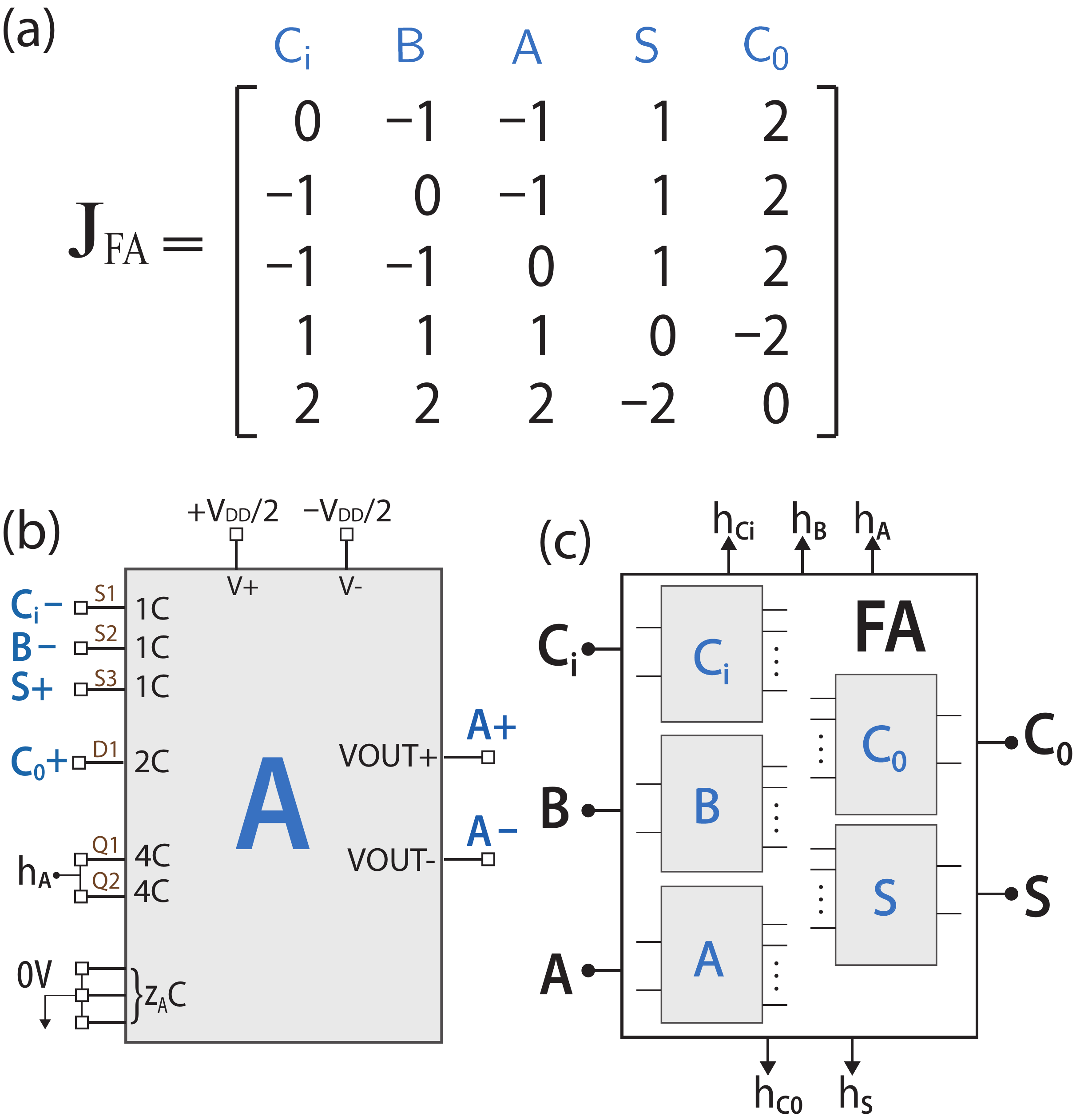}
 \caption{\textbf{Invertible Full Adder with $^W\hspace{-4pt}p$-bit}: (a)$[J]$ matrix for implementing a Full Adder. (b) Explicitly shows the hardware connections made to one of the input $p$-bits (A) from the other $p$-bits where 1$C$, 2$C$, and 4$C$ represent capacitors in units of $C=C_0=100aF$. (c) Shows the subcircuit representation of the Full Adder with its input/output terminals; $C_i,B,A$ input and $S,C_o$ output read terminals and separate corresponding clamping terminals $h_{C_i},h_B,h_A, h_S, h_{C_0}$. We used 8$C$ for the clamping terminals to ensure input / outputs follow what is dictated by the external signals. }
 \label{fi:fig2}
\end{figure}
%%%%%%%%%%%%%%%%%%%%%%%%%%

\begin{figure}[!t]
\centering
\includegraphics[width=0.97 \linewidth]{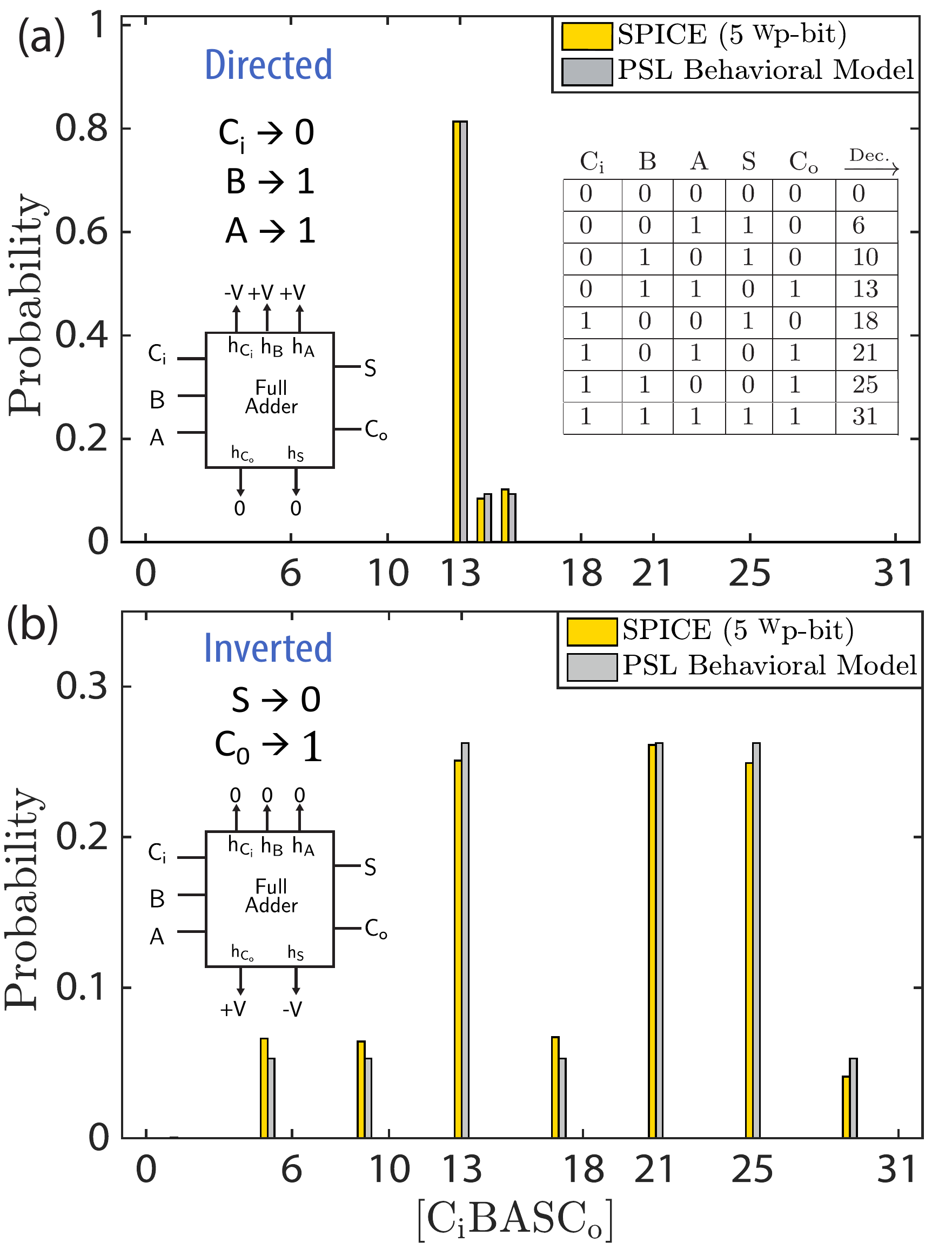}
\caption{\textbf{Full SPICE implementation of an Invertible Full Adder(5 $^W\hspace{-4pt}p$-bit)}: The 5 $^W\hspace{-4pt}p$-bit invertible Full Adder circuit is simulated in (a) Directed and (b) Inverted modes. The clamping values are indicated. All biasing terminals that are not clamped to 1 or 0 are grounded. The histogram of [$\rm{C_i B A S C_0}$] is obtained after thresholding voltages ($(V<0)\equiv-1, (V>0)\equiv+1$). The SPICE model is run for $1\rm{\mu s}$ and compared with the PSL equations where each $p$-bit is updated in random but sequential order \cite{camsari2017stochastic}. In this example $I_0\simeq 1$ is chosen to emphasize how the models are in good agreement even in the magnitudes of the minor peaks of the histogram.}
\label{fi:fig3}
\end{figure}

A Full Adder can be implemented in PSL using the $[J]$ matrix shown in Fig.~\ref{fi:fig2}. In this paper, we improve the 14 $p$-bit implementation of the invertible Full Adder (FA) in Ref.\cite{camsari2017stochastic} and implement the same functionality using 5 $p$-bits. This is achieved by first noting that the first half  of the FA truth table is complementary to the second half for the FA (Fig.~\ref{fi:fig3}a inset). The first 4 lines in the truth table is turned into an orthonormal set by a Gram-Schmidt process and a [J] matrix is obtained using Eq.12 in Ref.\cite{camsari2017stochastic} which is finally rounded to integer values, with diagonal entries replaced by zeros. This $[J]$ defines the interconnection between the 5 $^W\hspace{-4pt}p$-bits of the Full Adder in hardware. Each row of the $[J]$ matrix are realized in terms of capacitive coupling to the gate of the associated terminal.

%%%%%%%%%%%%%%%%%%%%%%%%%%%%%
%%%%%%%%%%%%%%%%%%%%%%%%%%%%%%

To ensure a uniform $I_0$ is applied to each $p$-bit (Eq.~\ref{eq:io}), the same weighting factor K  needs to be used for all $^W\hspace{-4pt}p$-bits.  To apply a given $I_0$, we first find  max($b_i+\sum J_{ij}$) for any given $[J]$, and then ground $z_i=M-b_i+ \sum J_{ij}$ ($z_i \ge 0, z_i \in N)$ unit capacitances for all terminals where $M$ is a number that can be used to control $I_0$, a larger $M$ causing a smaller $I_0$. Fig.~\ref{fi:fig2}b shows explicit connections made to one of the inputs ``A'' and Fig.~\ref{fi:fig2}c shows the subcircuit  of the Full Adder with $C_i, B, A$ as inputs, $S, C_0$ as the outputs, and $h_{Ci}, h_B, h_A, h_S, h_{Co}$ as the clamping pins.

%\textbf{Explain Results}
Fig.~\ref{fi:fig4} shows the operation of a Full Adder in the usual forward mode with $C_i, B, A$ clamped to values (0,1,1) which forces the $S~\rm{and}~C_0$ to (0,1) according to the truth table. In the invertible mode $S~\rm{and}~C_0$ are clamped to (0,1) and the circuit stochastically searches \emph{consistent} combinations of $C_i, B, A$ to satisfy the truth table: $\{C_i, B, A \} =\{\{0,1,1\},\{1,0,1\},\{1,1,0\}\}$. Fig.~\ref{fi:fig4} shows steady state (t = 1 $\mu s$) histogram plots of the Full Adder operation in direct and inverted mode side by side with results from the PSL behavioral model.

The good agreement between the ideal PSL behavioral model and the coupled SPICE simulation that solves PTM-based transistors models with stochastic LLG validates the hardware mapping of the ideal $p$-bit equations with the weighted $p$-bits.

\section{3SUM Problem}

%\textbf{Introduce 3SUM Problem}\\
3SUM is a decision problem in complexity theory that asks whether three elements of a given set can sum up to zero.  A variant of the problem is when the set of three numbers have to add up to a given constant number. This problem has a polynomial time solution and is not in NP.  In this section, we show how the invertibility feature of the Full Adders can be utilized to design a hardware 3SUM solver, and in the next section, we show how the 3SUM hardware can be modified to design a general solver for the NP-complete Subset Sum Problem. 

%%%%%%%%%%%%%%%%%%%%%%%%%%%%%%
\begin{figure}[!h]
\centering
\includegraphics[width=0.99\linewidth]{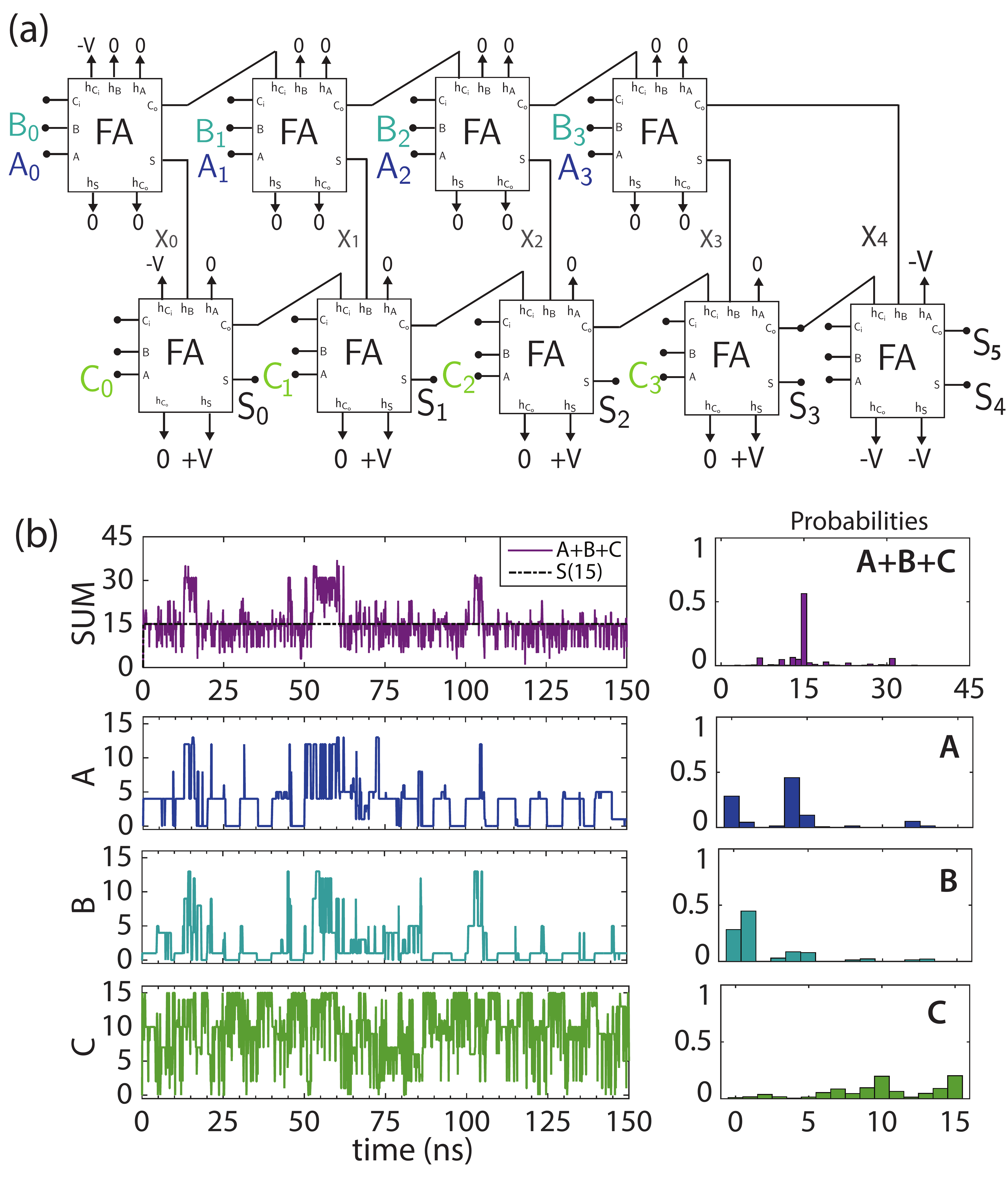}
 \caption{\textbf{SPICE simulation of a 4bit 3-SUM Problem (9 $\times$ 5 =  45 $^W\hspace{-4pt}p$-bit network)}: (a) The circuit is constructed by interconnecting two rows of invertible Full-Adders (FA) to construct a 3 number, 4-bit adder. The sum S is clamped to the desired value and A, B, C resolves themselves to create all the possible 3 number subsets out of all positive numbers 0 to $2^4-1$ that satisfy $\rm{A+B+C=S}$. (b) Shows the results when S is clamped to 15. A, B and C get correlated to satisfy the sum with different combinations. In this example, the inputs A, B, C are unconstrained and can take on any value between $0-15$.  }
\label{fi:fig4}
\end{figure}
%%%%%%%%%%%%%%%%%%%%%%%%%%%%%%

The invertibility property of the Full Adders ensure that given the sum, it can provide the possible input combinations for that sum as shown in Fig.\ref{fi:fig4}a. So an n-bit 3 number adder circuit implemented in PSL can essentially provide solution sets for the 3SUM problem when the sum is clamped to a given value.

Fig.~\ref{fi:fig4}a shows the circuit constructed out of Full Adders to solve a 4-bit 3SUM problem. Each of the Full Adders in the circuit are the 5 $p$-bit invertible adders that were shown in Fig.~\ref{fi:fig3}. The first row of adders adds the two 4-bit numbers A and B, and feeds its output X, to the next row of adders which adds X and C to give the sum $S=C+X=C+B+A$.  Because $p$-circuits are invertible, if we clamp the sum S, the circuit naturally explores through all possible sets and multisets of the set of all integers from 0 to $2^4-1$ that add up to S. The given set for the problem could be implemented through clamping certain bits of A,B and C or externally circuitry could be used to detect only the results that belong to the given set. Fig.~\ref{fi:fig4}b shows the how A,B,C is fluctuating between values that satisfy the clamped sum 15.

\section{Subset-sum Problem (SSP)} 

\begin{figure}[!t]
\centering
\includegraphics[width=0.99\linewidth]{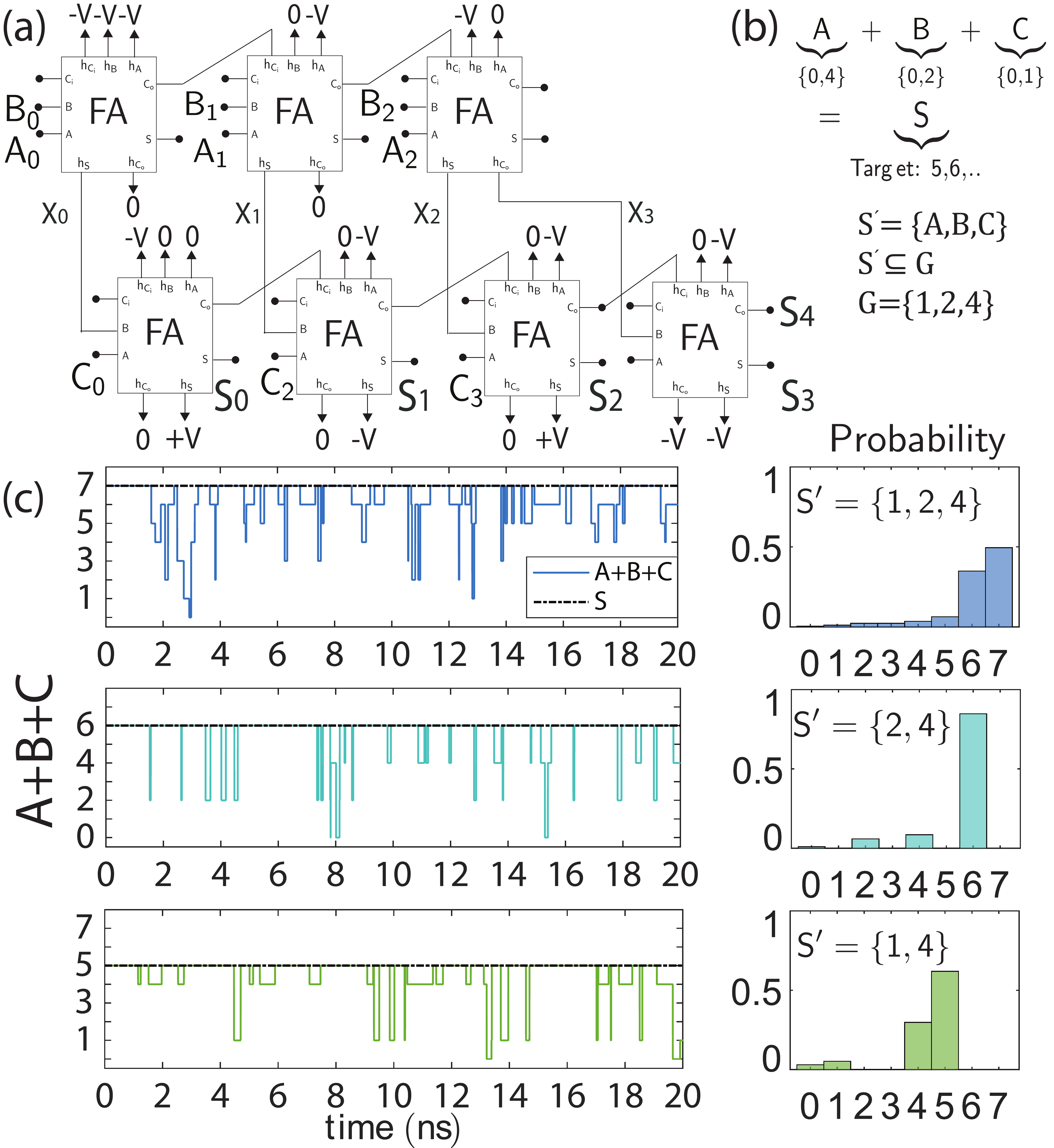}
 \caption{\textbf{SPICE simulation of a 3 input, 3-bit Subset Sum Problem (7 $\times$ 5 = 35 $^W\hspace{-4pt}p$-bit network)}: (a) A 3-input 3-bit binary adder that adds three numbers A,B,C. Unlike the 3SUM, in this case the inputs are constrained to a given value specified by the set G =$\{1,2,4\}$ in this example. A target S is selected and the output of the adders are clamped to the target value as shown in (b).  (c) Shows three different instances of a target where the inputs find a consistent combination (the correct subset of G) to satisfy the target.  Histograms show that the highest probable state is the correct subset. An important difference from the 3SUM circuit is that the information flow is \emph{directed} from the target (second layer of adders) to the first layer of adders.}
\label{fi:fig5}
\end{figure}

In this section, we show how the hardware circuit that was designed for 3SUM problem could be modified to solve a small instance of  subset-sum problem (SSP) \cite{cormen2009introduction} which is believed to be a fundamentally difficult  problem in computer science (NP-complete). The SSP asks, given a set G with a finite number of positive numbers, if there is a subset S' such that S' $\subseteq$ G whose elements sum to a specified target. For example, Fig.~\ref{fi:fig5} shows a circuit that is programmed to choose a set, G=$\{1,2,4\}$ and a target that is defined by 4-bits. In the 3SUM circuit the input bits (A, B, C) were left ``floating'',  here, the inputs are constrained to a given number (1,2,4) by clamping the remaining bits of an input. For example, the inputs $A_1$ and $A_0$ are clamped to zero to make A either 4 or 0. Under these conditions, clamping the output to a specified target makes the circuit search for a \emph{consistent} input combination to find a subset that satisfies the clamped target. Fig.~\ref{fi:fig5}c shows three example targets where the inputs get correlated to satisfy the clamped sum. The invertibility feature that is utilized to solve the SSP in this hardware is similar to those discussed in the context of memcomputing \cite{traversa2017polynomial}, however the physical mechanisms are completely different.

One striking difference in the design of the SSP we considered, compared to the 3SUM hardware is the \emph{direction} of information. In 3SUM the connections  were from the first layer of  Full Adders to the second, as in normal addition (Fig.~\ref{fi:fig4}a). In the SSP, we observed that reversing these connections from the second layer of adder to the first layer drastically improves the accuracy of the solution (Fig.~\ref{fi:fig5}a). A similar observation regarding the directional flow of information for another inverse problem using $p$-circuits (integer factorization) was made in \cite{camsari2017stochastic}.  Here we have limited the discussion to a small instance of the SSP which would in general require more layers of Full Adders in both vertical and horizontal directions to account for more numbers of elements in G and their size.  The purpose of this example is to illustrate how invertibility can be combined with standard digital VLSI design to construct any general ``cost function'' for hard problems of computer science in an asynchronously running \textit{hardware} platform without any external clocking.

\section{Conclusion}
In this paper we have proposed a compact building-block for Probabilistic Spin Logic (PSL) combining a recently proposed Embedded MRAM-based $p$-bit, with an integrated capacitive network that can be implemented using Floating Gate MOS (FGMOS) transistors similar to the neuMOS concept. We have shown by extensive SPICE simulations that the results of the hardware model for the weighted $p$-bit agree well with the behavioral equations of PSL. Having dedicated MTJ based hardware stochastic neurons could help minimize the footprint and consume lower power for applications as also indicated by ref.\cite{zand2017r,mizrahi2018neural}. Even though an FGMOS-based capacitive network for performing the voltage addition seems like a natural option, we note that the device equations for any capacitance [$\sf C_{ij}$] or conductance network [$\sf G_{ij}]$ would have been essentially the same. Moreover, our discussion was only about static weights, but an FPGA-like reconfigurable weighting scheme can also be employed either by using transistor-based gates or by additional multiplexing circuitry to perform online learning or to redesign $p$-circuit connectivity. Finally, using the basic building block we have shown how a small instance of the NP-complete Subset Sum Problem hardware solver can be designed using the unique invertibility feature of $p$-circuits. 

\section*{Acknowledgment}
This work was supported in part by the Center for Probabilistic Spin Logic for Low-Energy Boolean and Non-Boolean Computing  (CAPSL), one of the Nanoelectronic Computing Research (nCORE) Centers as task 2759.005, a Semiconductor Research Corporation (SRC) program sponsored by the NSF through ECCS 1739635.
%\bibliography{PSL8ref}

\begin{thebibliography}{12}%
\makeatletter
\providecommand \@ifxundefined [1]{%
 \@ifx{#1\undefined}
}%
\providecommand \@ifnum [1]{%
 \ifnum #1\expandafter \@firstoftwo
 \else \expandafter \@secondoftwo
 \fi
}%
\providecommand \@ifx [1]{%
 \ifx #1\expandafter \@firstoftwo
 \else \expandafter \@secondoftwo
 \fi
}%
\providecommand \natexlab [1]{#1}%
\providecommand \enquote  [1]{``#1''}%
\providecommand \bibnamefont  [1]{#1}%
\providecommand \bibfnamefont [1]{#1}%
\providecommand \citenamefont [1]{#1}%
\providecommand \href@noop [0]{\@secondoftwo}%
\providecommand \href [0]{\begingroup \@sanitize@url \@href}%
\providecommand \@href[1]{\@@startlink{#1}\@@href}%
\providecommand \@@href[1]{\endgroup#1\@@endlink}%
\providecommand \@sanitize@url [0]{\catcode `\\12\catcode `\$12\catcode
  `\&12\catcode `\#12\catcode `\^12\catcode `\_12\catcode `\%12\relax}%
\providecommand \@@startlink[1]{}%
\providecommand \@@endlink[0]{}%
\providecommand \url  [0]{\begingroup\@sanitize@url \@url }%
\providecommand \@url [1]{\endgroup\@href {#1}{\urlprefix }}%
\providecommand \urlprefix  [0]{URL }%
\providecommand \Eprint [0]{\href }%
\providecommand \doibase [0]{http://dx.doi.org/}%
\providecommand \selectlanguage [0]{\@gobble}%
\providecommand \bibinfo  [0]{\@secondoftwo}%
\providecommand \bibfield  [0]{\@secondoftwo}%
\providecommand \translation [1]{[#1]}%
\providecommand \BibitemOpen [0]{}%
\providecommand \bibitemStop [0]{}%
\providecommand \bibitemNoStop [0]{.\EOS\space}%
\providecommand \EOS [0]{\spacefactor3000\relax}%
\providecommand \BibitemShut  [1]{\csname bibitem#1\endcsname}%
\let\auto@bib@innerbib\@empty
%</preamble>
%%[1]
\bibitem [{\citenamefont {Sutton}\ \emph {et~al.}(2017)\citenamefont {Sutton},
  \citenamefont {Camsari}, \citenamefont {Behin-Aein},\ and\ \citenamefont
  {Datta}}]{sutton2017intrinsic}%
    \BibitemOpen
  \bibfield  {author} {\bibinfo {author} {\bibfnamefont {B.}~\bibnamefont
  {Sutton}}, \bibinfo {author} {\bibfnamefont {K.~Y.}\ \bibnamefont {Camsari}},
  \bibinfo {author} {\bibfnamefont {B.}~\bibnamefont {Behin-Aein}}, \ and\
  \bibinfo {author} {\bibfnamefont {S.}~\bibnamefont {Datta}},\ }\href
  {https://doi.org/10.1038/srep44370} {\bibfield  {journal} {\bibinfo
  {journal} {Scientific Reports}\ }\textbf {\bibinfo {volume} {7}} (\bibinfo
  {year} {2017})}\BibitemShut {NoStop}%
%%[2]  
\bibitem [{\citenamefont {Behin-Aein}\ \emph {et~al.}(2016)\citenamefont
  {Behin-Aein}, \citenamefont {Diep},\ and\ \citenamefont
  {Datta}}]{behin2016building}%
  \BibitemOpen
  \bibfield  {author} {\bibinfo {author} {\bibfnamefont {B.}~\bibnamefont
  {Behin-Aein}}, \bibinfo {author} {\bibfnamefont {V.}~\bibnamefont {Diep}}, \
  and\ \bibinfo {author} {\bibfnamefont {S.}~\bibnamefont {Datta}},\ }\href
  {http://dx.doi.org/10.1038/srep29893} {\bibfield  {journal} {\bibinfo
  {journal} {Scientific reports}\ }\textbf {\bibinfo {volume} {6}} (\bibinfo
  {year} {2016})}\BibitemShut {NoStop}%  
%%[3]  
\bibitem [{\citenamefont {Shim}\ \emph {et~al.}(2017)\citenamefont {Shim},
  \citenamefont {Jaiswal},\ and\ \citenamefont {Roy}}]{shim2017}%
  \BibitemOpen
  \bibfield  {author} {\bibinfo {author} {\bibfnamefont {Y.}~\bibnamefont
  {Shim}}, \bibinfo {author} {\bibfnamefont {A.}~\bibnamefont {Jaiswal}}, \
  and\ \bibinfo {author} {\bibfnamefont {K.}~\bibnamefont {Roy}},\ }\href
  {\doibase 10.1063/1.4983636} {\bibfield  {journal} {\bibinfo  {journal}
  {Journal of Applied Physics}\ }\textbf {\bibinfo {volume} {121}},\ \bibinfo
  {pages} {193902} (\bibinfo {year} {2017})},\ \Eprint
  {http://arxiv.org/abs/http://dx.doi.org/10.1063/1.4983636}
  {http://dx.doi.org/10.1063/1.4983636} \BibitemShut {NoStop}%
%%[4]  
\bibitem [{\citenamefont {Camsari}\ \emph
  {et~al.}(2017{\natexlab{a}})\citenamefont {Camsari}, \citenamefont {Faria},
  \citenamefont {Sutton},\ and\ \citenamefont {Datta}}]{camsari2017stochastic}%
  \BibitemOpen
  \bibfield  {author} {\bibinfo {author} {\bibfnamefont {K.~Y.}\ \bibnamefont
  {Camsari}}, \bibinfo {author} {\bibfnamefont {R.}~\bibnamefont {Faria}},
  \bibinfo {author} {\bibfnamefont {B.~M.}\ \bibnamefont {Sutton}}, \ and\
  \bibinfo {author} {\bibfnamefont {S.}~\bibnamefont {Datta}},\ }\href
  {\doibase 10.1103/PhysRevX.7.031014} {\bibfield  {journal} {\bibinfo
  {journal} {Phys. Rev. X}\ }\textbf {\bibinfo {volume} {7}},\ \bibinfo {pages}
  {031014} (\bibinfo {year} {2017}{\natexlab{a}})}\BibitemShut {NoStop}%  
%%[5]  
\bibitem [{\citenamefont {Zand}\ \emph {et~al.}(2017)\citenamefont {Zand},
  \citenamefont {Camsari}, \citenamefont {Ahmed}, \citenamefont {Pyle},
  \citenamefont {Kim}, \citenamefont {Datta},\ and\ \citenamefont
  {DeMara}}]{zand2017r}%
  \BibitemOpen
  \bibfield  {author} {\bibinfo {author} {\bibfnamefont {R.}~\bibnamefont
  {Zand}}, \bibinfo {author} {\bibfnamefont {K.~Y.}\ \bibnamefont {Camsari}},
  \bibinfo {author} {\bibfnamefont {I.}~\bibnamefont {Ahmed}}, \bibinfo
  {author} {\bibfnamefont {S.~D.}\ \bibnamefont {Pyle}}, \bibinfo {author}
  {\bibfnamefont {C.~H.}\ \bibnamefont {Kim}}, \bibinfo {author} {\bibfnamefont
  {S.}~\bibnamefont {Datta}}, \ and\ \bibinfo {author} {\bibfnamefont {R.~F.}\
  \bibnamefont {DeMara}},\ }\href@noop {} {\bibfield  {journal} {\bibinfo
  {journal} {arXiv preprint arXiv:1710.00249}\ } (\bibinfo {year}
  {2017})}\BibitemShut {NoStop}%  
%%[6]  
\bibitem [{\citenamefont {Lin}\ \emph {et~al.}(2009)\citenamefont {Lin},
  \citenamefont {Kang}, \citenamefont {Wang}, \citenamefont {Lee},
  \citenamefont {Zhu}, \citenamefont {Chen}, \citenamefont {Li}, \citenamefont
  {Hsu}, \citenamefont {Kao}, \citenamefont {Liu} \emph
  {et~al.}}]{lin200945nm}%
  \BibitemOpen
  \bibfield  {author} {\bibinfo {author} {\bibfnamefont {C.}~\bibnamefont
  {Lin}}, \bibinfo {author} {\bibfnamefont {S.}~\bibnamefont {Kang}}, \bibinfo
  {author} {\bibfnamefont {Y.}~\bibnamefont {Wang}}, \bibinfo {author}
  {\bibfnamefont {K.}~\bibnamefont {Lee}}, \bibinfo {author} {\bibfnamefont
  {X.}~\bibnamefont {Zhu}}, \bibinfo {author} {\bibfnamefont {W.}~\bibnamefont
  {Chen}}, \bibinfo {author} {\bibfnamefont {X.}~\bibnamefont {Li}}, \bibinfo
  {author} {\bibfnamefont {W.}~\bibnamefont {Hsu}}, \bibinfo {author}
  {\bibfnamefont {Y.}~\bibnamefont {Kao}}, \bibinfo {author} {\bibfnamefont
  {M.}~\bibnamefont {Liu}},  \emph {et~al.},\ }\ \href
  {https://doi.org/10.1109/IEDM.2009.5424368} {\emph {\bibinfo {booktitle}
  {Electron Devices Meeting (IEDM), 2009 IEEE International}}}\ (\bibinfo
  {organization} {IEEE},\ \bibinfo {year} {2009})\ pp.\ \bibinfo {pages}
  {1--4}\BibitemShut {NoStop}%  
%%[7]  
\bibitem [{\citenamefont {Camsari}\ \emph
  {et~al.}(2017{\natexlab{b}})\citenamefont {Camsari}, \citenamefont
  {Salahuddin},\ and\ \citenamefont {Datta}}]{camsari2017implementing}%
  \BibitemOpen
  \bibfield  {author} {\bibinfo {author} {\bibfnamefont {K.~Y.}\ \bibnamefont
  {Camsari}}, \bibinfo {author} {\bibfnamefont {S.}~\bibnamefont {Salahuddin}},
  \ and\ \bibinfo {author} {\bibfnamefont {S.}~\bibnamefont {Datta}},\
  }\href@noop {} {\bibfield  {journal} {\bibinfo  {journal} {IEEE Electron
  Device Letters}\ }\textbf {\bibinfo {volume} {38}},\ \bibinfo {pages} {1767}
  (\bibinfo {year} {2017}{\natexlab{b}})}\BibitemShut {NoStop}%
%%[8]  
\bibitem [{\citenamefont {Shibata}\ and\ \citenamefont
  {Ohmi}(1992)}]{ohmi1992neumos}%
  \BibitemOpen
  \bibfield  {author} {\bibinfo {author} {\bibfnamefont {T.}~\bibnamefont
  {Shibata}}\ and\ \bibinfo {author} {\bibfnamefont {T.}~\bibnamefont {Ohmi}},\
  }\href@noop {} {\bibfield  {journal} {\bibinfo  {journal} {IEEE Transactions
  on Electron devices}\ }\textbf {\bibinfo {volume} {39}},\ \bibinfo {pages}
  {1444} (\bibinfo {year} {1992})}\BibitemShut {NoStop}%
%%[9]  
\bibitem [{\citenamefont {Mizrahi}(2018)}]{mizrahi2018neural}%
  \BibitemOpen
  \bibfield  {author} {\bibinfo {author} {\bibfnamefont {A.}\ \bibnamefont
  {Mizrahi}}, \bibinfo {author} {\bibfnamefont {T.}~\bibnamefont
  {Hirtzlin}}, \bibinfo {author} {\bibfnamefont {A.}~\bibnamefont
  {Fukushima}}, \bibinfo {author} {\bibfnamefont {K.}~\bibnamefont
  {Hitoshi}}, \bibinfo {author} {\bibfnamefont {Y.} \bibnamefont
  {Shinji}}, \bibinfo {author} {\bibfnamefont {J.} \bibnamefont
  {Grollier}}\ and\ \bibinfo {author} {\bibfnamefont {D.}~\bibnamefont
  {Querlioz}},\ }\href@noop {} {\emph {\bibinfo {title} {Neural-like computing with populations of superparamagnetic basis functions}}}\ \bibinfo  {publisher} {Nature communications},\textbf {\bibinfo {volume} {9}},\ \bibinfo {pages}{1533} \ (\bibinfo {year}{2018})\BibitemShut {NoStop}%  
%%[10]  
\bibitem [{\citenamefont {Nakamura}\ \emph {et~al.}(2015)\citenamefont
  {Nakamura}, \citenamefont {Shimada}, \citenamefont {Matsuda},\ and\
  \citenamefont {Kimura}}]{nakamura2015neuron}%
  \BibitemOpen
  \bibfield  {author} {\bibinfo {author} {\bibfnamefont {N.}~\bibnamefont
  {Nakamura}}, \bibinfo {author} {\bibfnamefont {K.}~\bibnamefont {Shimada}},
  \bibinfo {author} {\bibfnamefont {T.}~\bibnamefont {Matsuda}}, \ and\
  \bibinfo {author} {\bibfnamefont {M.}~\bibnamefont {Kimura}},\ }\
  \href@noop {} {\emph {\bibinfo {booktitle} {Future of Electron Devices,
  Kansai (IMFEDK), 2015 IEEE International Meeting for}}}\ (\bibinfo
  {organization} {IEEE},\ \bibinfo {year} {2015})\ pp.\ \bibinfo {pages}
  {90--91}\BibitemShut{NoStop}%
%%[11]  
\bibitem [{\citenamefont {Cormen}(2009)}]{cormen2009introduction}%
  \BibitemOpen
  \bibfield  {author} {\bibinfo {author} {\bibfnamefont {T.~H.}\ \bibnamefont
  {Cormen}},\ }\href@noop {} {\emph {\bibinfo {title} {Introduction to
  algorithms}}}\ (\bibinfo  {publisher} {MIT press},\ \bibinfo {year}
  {2009})\BibitemShut{NoStop} %
%%[12]  
\bibitem [{\citenamefont {Traversa}\ and\ \citenamefont
  {Di~Ventra}(2017)}]{traversa2017polynomial}%
  \BibitemOpen
  \bibfield  {author} {\bibinfo {author} {\bibfnamefont {F.~L.}\ \bibnamefont
  {Traversa}}\ and\ \bibinfo {author} {\bibfnamefont {M.}~\bibnamefont
  {Di~Ventra}},\ }\href {https://doi.org/10.1063/1.4975761} {\bibfield
  {journal} {\bibinfo  {journal} {Chaos: An Interdisciplinary Journal of
  Nonlinear Science}\ }\textbf {\bibinfo {volume} {27}},\ \bibinfo {pages}
  {023107} (\bibinfo {year} {2017})}\BibitemShut {NoStop}%
\end{thebibliography}

%%%[4]  
%\bibitem [{\citenamefont {Faria}\ \emph {et~al.}(2018)\citenamefont {Faria},
%  \citenamefont {Camsari},\ and\ \citenamefont
%  {Datta}}]{faria2018implementing}%
%  \BibitemOpen
%  \bibfield  {author} {\bibinfo {author} {\bibfnamefont {R.}~\bibnamefont
%  {Faria}}, \bibinfo {author} {\bibfnamefont {K.~Y.}\ \bibnamefont {Camsari}},
%  \ and\ \bibinfo {author} {\bibfnamefont {S.}~\bibnamefont {Datta}},\
%  }\href@noop {} {\bibfield  {journal} {\bibinfo  {journal} {arXiv preprint
%  arXiv:1801.00497}\ } (\bibinfo {year} {2018})}\BibitemShut {NoStop}%  

 \end{document}